# Dynamic quantum sensing of paramagnetic species using nitrogen-vacancy centers in diamond

Valentin Radu[1,∇], Joshua Colm Price[1,∇], Simon James Levett[1], Kaarjel K. Narayanasamy[2], Thomas David Bateman-Price[1], Philippe Barrie Wilson[3], Melissa Louise Mather[1,*].

[1]Optics and Photonics Research Group, Faculty of Engineering, University Park, University of Nottingham, Nottingham, NG7 2RD, UK.

[2]School of Biomedical Sciences, University of Leeds, Leeds LS2 9JT, UK.

[3]Leicester School of Pharmacy, De Montfort University, The Gateway, Leicester LE1 9BH, UK).

[*]Corresponding Author

∇ These authors contributed equally.

**ABSTRACT:** Naturally occurring paramagnetic species (PS), such as free radicals and paramagnetic metalloproteins, play an essential role in a multitude of critical physiological processes including metabolism, cell signaling and immune response. These highly dynamic species can also act as intrinsic biomarkers for a variety of disease states whilst synthetic paramagnetic probes targeted to specific sites on biomolecules enable the study of functional information such as tissue oxygenation and redox status in living systems. The work presented herein describes a new sensing method that exploits the spin dependent emission of photoluminescence (PL) from an ensemble of nitrogen vacancy centers in diamond for rapid, non-destructive detection of PS in living systems. Uniquely this approach involves simple measurement protocols that assess PL contrast with and without the application of microwaves. The method is demonstrated to detect concentrations of paramagnetic salts in solution and the widely used magnetic resonance imaging contrast agent Gadobutrol with a limit of detection of less than 10 attomol over a 100 μm x 100 μm field of view. Real time monitoring of changes in the concentration of paramagnetic salts is demonstrated with image exposure times of 20 ms. Further, dynamic tracking of chemical reactions is demonstrated via the conversion of low spin cyanide coordinated $Fe^{3+}$ to hexaaqua $Fe^{3+}$ under acidic conditions. Finally, the capability to map paramagnetic species in model cells with sub-cellular resolution is demonstrated using lipid membranes containing gadolinium labelled phospholipids under ambient conditions in the order of minutes. Overall, this work introduces a new sensing approach for the realization of fast, sensitive imaging of PS in a widefield format that is readily deployable in biomedical settings. Ultimately this new approach to NV based quantum sensing paves the way towards minimally invasive real-time mapping and observation of free radicals in in vitro cellular environments.

Paramagnetic species (PS), which contain at least one unpaired electron in their valence shell, play critical roles both in normal physiology and in the pathophysiology of many diseases.[1–4] Free radicals, including superoxide, hydroxyl, reactive oxygen species and reactive nitrogen species, are important endogenous PS involved in numerous physiological processes including the immune response to infection[5], cell signaling[3] and redox regulation.[6] Moreover, elevated levels of free radicals can alter structural and functional properties of vital molecules including lipids, proteins, and nucleic acids, leading to extensive tissue dysfunction and injury.[7] Further, several endogenous PS have potential as biomarkers for a variety of disease states.[1–4] $Mn^{2+}$, for instance, has been related to developmental abnormalities in bone and skin[8,9] and a number of neurodegenerative diseases.[10] Similarly, variations in the bioavailability of the reactive nitrogen species nitric oxide is reportedly related to disease states such as pulmonary hypertension, diabetes, or heart failure.[11] At the same time, there is an expanding range of exogenous PS being used to probe function and molecular structure in living systems, via site specific targeting of biomolecules[12–15], and to extend the detection lifetime of short-lived endogenous PS via spin trapping.[16–18] Due to the multitude of roles PS play in biology there is a significant need for suitable sensing technologies capable of rapid detection of biologically relevant concentrations of PS under measurement conditions compatible with sustaining living systems.

Numerous approaches exist for detecting PS, however many of these are not satisfactory or suitable for use in living systems. Existing approaches include time consuming colorimetric assays addressing non-specific by-products[19–21] or biologically incompatible methods such as magnetic resonance force microscopy[22] or superconducting quantum interference devices (SQUIDS)[23] which require sample preparation and data acquisition at cryogenic temperatures. Furthermore, fluorescent probes which are widely established as probes for monitoring redox states in live cells, have numerous shortfalls such as cytotoxicity, photobleaching and nonspecific interactions often leading to inaccurate results.[21,24] Of the available approaches Electron Paramagnetic Resonance (EPR) spectroscopy is arguably one of the most powerful and established methods for detection of PS in biological samples. EPR spectroscopy allows for the direct detection of long-lived PS, as well as short-lived PS when functionalized with spin traps or spin probes. The technique also offers high chemical specificity and the ability to monitor changes in oxidation state.[7] EPR spectroscopy however has inherently low sensitivity demanding high concentrations or large sample volumes, neither of which are practical for biological samples. Further, important endogenous PS are typically produced in intracellular compartments such as the mitochondria and as such mapping these species require microscale imaging capabilities, which is orders of magnitude beyond current EPR imaging capabilities.[25,26] Thus, alternative strategies are required to attain the combination of spatio-temporal resolution, spin sensitivity and the operability under ambient conditions to detect PS in cellular environments.

The Nitrogen-Vacancy (NV) color center in diamond has emerged as a high performance quantum sensor capable of detecting PS. The NV color center is a naturally occurring paramagnetic impurity comprising of a substitutional Nitrogen atom adjacent to a vacant lattice site, which in the negative charge state (NV-) forms a spin triplet ground state energy level having $|m_s = 0\rangle$ and degenerate $|m_s = \pm 1\rangle$ spin states. Optical pumping of the $|m_s = 0\rangle$ ground state with green light results in a broadband PL emission with a zero phonon line at 637 nm and longer wavelength emission extending into the infrared region. NV- centers display remarkable chemical stability and photo-stability and have been used for a wide range of applications in biological sciences as fluorescent imaging probes.[27–30] Key to the NV- sensing capabilities is the addition of microwave excitation, which can be employed to manipulate the ground and excited state spin levels such that the PL intensity of the optically pumped system reduces when electrons occupy the $|m_s = \pm 1\rangle$ spin states. The degeneracy of the $|m_s = \pm 1\rangle$ spin states can be lifted by the application of external magnetic fields. The lifting of the degeneracy enables a variety of highly sensitive sensing protocols which in recent years have led to the application of the NV- center as a sensor for measuring changes in external perturbations including magnetic and electric fields[31–36], temperature[27,37,38], pressure and lattice strain.[39,40]

Amongst the catalogue of NV based sensing protocols is longitudinal spin (T1) based relaxometry, whereby rapidly fluctuating magnetic fields are quantified based on the rate of mixing of spin states in the ground state triplet after optical polarization into the ground state $|m_s = 0\rangle$. This technique has been successfully implemented for detection of various PS, such as $Gd^{3+}$ [41,42], $Mn^{2+}$ and ferritin.[43,44] Whilst these methods prove highly sensitive, and can provide spatial as well as quantitative information, they require the application of carefully constructed laser and microwave pulse sequences and measurement times in the order of seconds to hours to achieve an accurate readout of concentration, together with an optimal signal to noise ratio (SNR) necessary for high resolution quantitative imaging.

The work reported herein presents a highly sensitive, minimally invasive dynamic sensing technique based on the contrast of PL emission from an ensemble of shallow NV- centers in diamond with and without microwaves. The technique is demonstrated and limit of detection determined using aqueous solutions of gadolinium nitrate ($Gd(NO_3)_3$), ferric chloride ($FeCl_3$) and a widely used Magnetic Resonance Imaging (MRI) contrast agent Gadobutrol. The ability to dynamically track changes in spin state is also demonstrated via conversion of low spin cyanide coordinated $Fe^{3+}$ to hexaaqua $Fe^{3+}$ under acidic conditions. The capability to image PS on a subcellular level is also demonstrated using multi-lamellar liposomes containing $Gd^{3+}$-labelled phospholipids. Importantly the method described is implemented on a commercially available fluorescence microscope using the microscope control software and as such could be readily used by non-specialists in the biological community. Overall, the combination of spatio-temporal resolution, spin sensitivity and the operability under ambient conditions using widefield quantum sensing paves the way towards minimally invasive real-time observation of free radicals within in vitro cellular environments.

MATERIALS AND METHODS

Diamond samples

The diamond samples used in this study were 2.0 mm x 2.0 mm x 0.5 mm polished, electronic grade single-crystal diamond plates grown by chemical vapor deposition, with bulk nitrogen content [N] < 5 ppb and (100)-oriented top surface polished with < 0.5 nm Ra (purchased from Element Six, Ascot, UK). The diamond plates were implanted with $^{14}N^+$ and $^{15}N^+$ ions at an energy of 6 keV with a dose of $10^{13}$ ions/cm². Concentration profiles of the implanted $^{14}N^+$ ions were determined with the Monte Carlo simulation code Stopping and Range of Ions in Matter (SRIM) [http://www.srim.org] with a projected depth of 14.7 nm and straggle of 5.6 nm. Following implantation the samples were annealed under Nitrogen gas according to methods reported by Chu et al.[45] Prior to measurements the plates were cleaned by keeping them under acid reflux for 2 hours using a 1:1 mixture of concentrated sulfuric and nitric acid. This was followed by surface oxidation under air for 10 hours at 465 °C and another step of refluxing in acid mixture for 2 hours. The diamond plates were then rinsed with deionized water and kept in dichloromethane or dimethyl sulfoxide.

NV based sensing

The NV based sensing regime involved detection of PL from an ensemble of shallow NV centers in the diamond plates with and without the application of MWs. The experimental setup (Figure 1) involved mounting a diamond plate on a glass cover slip attached to a custom designed printed circuit board (PCB). The PCB was designed to have an aperture enabling illumination of the sample via an inverted fluorescence microscope (Olympus IX83) and electrical connection for delivery of MWs. A 0.125 mm diameter straight copper wire running across the top of the diamond plate and electrically connected to 50 Ohm tracks on the PCB was used to deliver MWs from a Keysight N5181B vector signal generator and an AR 20S1G4 MW amplifier. NV centers were illuminated with a mercury arc lamp filtered through a narrow band pass excitation filter centered at 532 nm. The light was subsequently focused on to the back focal plane of an oil-immersion 60x TIRF objective lens (NA = 1.49), providing a power density of 1 W/cm$^2$ at the sample position. Emitted PL was filtered through a 575 nm long pass filter and imaged onto a sCMOS camera (Photometrics 95 Prime B) providing a maximum field of view of 200 µm x 200 µm.

The detection method presented here is based on PS induced changes in PL contrast. Experimentally, the presence of PS was first performed via acquisition of optically detected magnetic resonance (ODMR) spectra using a continuous wave (CW) measurement regime in which illumination was constant and images of PL were acquired as the MW frequency was swept to probe separately the ground state (2.77 GHz to 2.97 GHz) and excited state (1.35 GHz to 1.55 GHz) NV spin transitions. Image acquisition was synchronized with MW sweeps using an Olympus Real-Time Controller (RTC) via the Olympus CellSens software (Tokyo, Japan). The resulting images at each frequency were analyzed to extract the PL intensity, determined by summing pixel intensities from 100 µm x 100 µm sized fields of view, over which the lamp intensity was uniform. At each frequency 10 repeats of measurements were performed for 3 different fields of view. ODMR spectra were produced by plotting average PL intensity from all repeats and fields of view against MW frequency. ODMR spectra were then normalized by dividing all average PL intensities by the maximum averaged intensity within the sweep.

Next a fast and simple measurement protocols was demonstrated that involved acquisition of images of PL with MWs on resonance (2.868 GHz) and MWs off. Exposure times of 20 ms were used for each image and PL intensity determined by summing pixel values across 100 µm x 100 µm sized fields of view in each image. The ratio of PL intensity for MWs on and MWs off was calculated and normalized by dividing this ratio by the PL intensity obtained for deionized water with MWs on and MWs off. Calibration curves plotting the normalized ratio of PL with MWs on to MWs off were produced using datasets consisting of 100 pairs of images. In this instance the PL intensity values were summed for each image and the average PL intensity determined.

Exemplar samples were used to study the effect of PS on ODMR spectra and contrast with MWs on and MWs off. Here, aqueous solutions of Gd(NO3)3, FeCl3, LaCl3 and Gadobutrol were chosen due to their varied paramagnetic properties and relevance to biological systems and bio-imaging. Indeed, Gd$^{3+}$ is of particular interest as it is strongly paramagnetic due to its 7 unpaired electrons in the 4f subshell. Here it is studied as a salt and as a chelate in the form of the widely used NMR contrast agent Gadobutrol. Fe$^{3+}$ was also chosen due to its 5 unpaired electrons and its relevance to a multitude of biological processes reliant on electron transfer.[46] Lanthanum(III) Chloride (LaCl$_3$) was used as a control as it is chemically similar to Gd$^{3+}$ however does not have any unpaired electrons and thus enables the assessment of the role paramagnetism plays in the NV sensing protocols described.

Each exemplar sample was separately pipetted (100 µL) with increasing concentration on to the diamond plates. Plates were washed with deionized water in between each concentration increase and acid cleaned between each solution type. The ability to dynamically track changes in the concentration of PS in solution was demonstrated using the MWon:MWoff protocol using temporally varying concentrations of aqueous solutions of Gd(NO3)3 and tracking the conversion of low spin cyanide coordinated Fe$^{3+}$ to hexaaqua Fe$^{3+}$ under acidic conditions. Here image exposure times as short as 20 ms were used.

Liposome preparation and imaging

Liposomes were used as cell models to demonstrate the potential of the MWon:MWoff protocol to map PS on a subcellular level. Liposomes were prepared at room temperature using HPLC grade chloroform, methanol, and buffer solution pH 7.4 (20 mM 4-Morpholinepropanesulfonic acid, 30 mM sodium sulfate). Chloroform-solubilized phosphatidylcholine (POPC, Avanti Polar Lipids) was mixed with biotinylated phosphatidylethanolamine (Biotinyl Cap PE) at a molar ratio of 10:1. Powdered Gd$^{3+}$-labelled phosphatidylethanolamine (PE-DTPA-Gd$^{3+}$, Avanti Polar Lipids) was re-suspended in a chloroform:methanol mixture (3:1 v/v) and added to the POPC:Biotinyl Cap PE mixture in chloroform to achieve a molar ratio of POPC:PE-DTPA-Gd$^{3+}$ of 6:1. The resulting mixture was dried under Nitrogen at room temperature and re-suspended in buffer solution. Control liposomes were prepared by mixing POPC and Biotinyl Cap PE at a molar ratio of 13:1.

Attachment of liposomes on the diamond plate involved first immersing the plate in an aqueous solution of 50 mM MES (2-(N-Morpholino)ethanesulfonic acid). EDC (N-(3-Dimethylaminopropyl)-N′-ethylcarbodiimide hydrochloride) and NHS (N-Hydroxysuccinimide) were added at a concentration of 52 mM and 87 mM, respectively. After fifteen minutes the plate was washed with buffer solution and immersed in a 4 µM streptavidin solution. After ten minutes the plate was washed with buffer solution and immersed in the 10-fold diluted liposome suspension for one hour. The plate was then washed and stored in buffer solution. Widefield images of the Gd containing liposomes and control liposomes attached to the diamond plates were subsequently acquired using the MWon:MWoff protocol. 5000 pairs of images with MWs on resonance and MWs off were obtained. Each stack of 5000 images obtained with MWs on and MWs off were summed and the ratio of the final summed images determined. Differential interference contrast (DIC) images were also recorded as a sum of a stack of 100 images to demonstrate liposome morphology and cross reference with the MWon:MWoff images.

RESULTS AND DISCUSSION

ODMR frequency sweeps were performed for exemplar solutions over a range of concentrations from 10 nM to 2 M. Concentrations above 1 M $Fe^{3+}$ were not tested as they are above the solubility limit. For clarity, Figure 2 displays ODMR spectra for selected concentrations between 100 nM and 1 M (see Supplementary Information Figure 1 for all concentrations). The results are indicative of ODMR spectra obtained from NV-color centers in that there is a reduction in PL around the ground state resonance frequency (2.87 GHz) for the transition between the $|m_s = 0\rangle$ and $|m_s = \pm 1\rangle$ spin state in the absence of an externally applied magnetic field. Near the resonance frequency a split is seen in all spectra reflective of intrinsic strain in the diamond chip.[47] Inspection of all plots shown in Figure 2 reveals the dependence of PL on the concentration and chemical form of PS. Indeed, as concentration of PS increases so too does the ratio between PL intensity on resonance to the highest PL intensity off resonance. This is particularly apparent for solutions containing $Gd^{3+}$ (both $Gd(NO_3)_3$ and Gadobutrol). The change in normalized PL intensity with concentration is less for $FeCl_3$ solutions and negligible for $LaCl_3$ solutions across the concentration range studied. Differences are also observed between the chelated form of $Gd^{3+}$ and free $Gd^{3+}$ which are known to differ in paramagnetic strength. Collectively these results point to a strong dependence of PL contrast on paramagnetic strength.

For all exemplar solutions insignificant changes in the linewidth (Supplementary Figure 2) and location of the resonance dip was observed, suggesting these parameters have a negligible effect on the contrast observed. Further, the change in normalized PL intensity was minimal with solution pH, the variation being within 1% across a pH range of 0.6-13, which is equivalent to the change induced by 1.1 µM $Gd^{3+}$ (Supplementary Table 1). Further measurements were made to investigate the ratio between PL intensity on resonance to the highest PL intensity off resonance around the excited state resonance (Supplementary Figure 3), which was also found to depend on PS concentration and paramagnetic strength as observed for the ground state transition. Collectively these results point to a dependence of the observed contrast on the concentration and paramagnetic strength of the solvated metal ions.

Figure 3 displays results obtained by measuring PL intensity with MW turned off and MW on resonance (2.868 GHz) with contrast being calculated as the ratio of PL intensity with MW on resonance to that with MW turned off. Normalized contrast (ratio of test solution normalized contrast to normalized contrast obtained for deionized water) is shown in the graphs and depends strongly on concentration for the two paramagnetic metal ions, namely $Gd^{3+}$ and $Fe^{3+}$ whilst negligible change is measured for $La^{3+}$. $Gd^{3+}$, which has a higher paramagnetic strength than $Fe^{3+}$, displays a stronger concentration dependence and is detected at lower concentration as compared to $Fe^{3+}$. The results depicted on a logarithmic concentration scale reveal an apparent plateau in $Gd(NO_3)_3$ results between 100 µM and 10 mM. It is hypothesized that this is related to the non-uniform distribution of ions in the detection volume which extends approximately 10 nm above the surface of the diamond chip. A similar effect was previously observed with aqueous $Mn^{2+}$ and was attributed to the inhomogeneous adsorption of ions on the diamond surface.[44] The non-uniform packing of the ions at the interface would affect the exchange rate of water molecules in the outer coordination sphere which significantly affects the spin properties of $Gd^{3+}$ ions.[48–51] The effect is not observed with chelated $Gd^{3+}$ as evidenced by the titration data recorded for titrations of Gadobutrol (Figures 3c and 3d) which exhibits a linear concentration dependency of the normalized contrast throughout the entire concentration range studied. Differences in the absolute values of contrast between the chelated $Gd^{3+}$ and free Gd are also seen for concentrations above 0.1 M with Gadobutrol producing contrast greater than 1.04 compared to lower contrast observed for free $Gd^{3+}$ (1.02 to 1.04). These results demonstrate the high sensitivity of the presented sensing regime not only to PS concentration but also to the coordination environment.

The capability of the sensing protocol to respond dynamically to changes in the concentration of PS was investigated by sequentially adding aqueous solutions with increasing concentration of $Gd(NO_3)_3$ to the surface of the diamond plate and monitoring contrast in PL intensity over time. Figure 4 displays the contrast obtained from the ratio of PL intensity with MW on (2.868 GHz) to that with MW off over time. Abrupt changes in contrast are observed corresponding to the addition of aliquots of solutions of higher concentration. These results serve as a demonstration that the simple measurement protocol presented allows fast detection of variations in the concentration of PS, which is of particular relevance for monitoring in biological systems. The time resolution of the measurement protocol is dependent on the acquisition time of a pair of images (MW on, MW off) and the SNR. The current protocol uses 20 ms exposure time for each image and 23.5 ms switching time (required by the control unit triggering the MW signal generator), therefore each data point in Figure 4 is acquired in 87 ms. Using the Nyquist criteria this equates to a maximum sampling rate of 5.7 Hz. There is scope for tremendous improvement in the signal to noise ratio and switching time. Lower exposure times could be readily deployed through detection of more photons (e.g. improved collection efficiency of emitted photons, improved quantum efficiency of the camera and increased light intensity). Such improvements will enable faster camera rates and in principle with the camera used in this work 1 kHz acquisition speed is possible corresponding to a maximum sampling rate of 250 Hz (4 ms) based on the Nyquist criteria and requirement of two images per data point.

The sensing protocol was also used to study the kinetics of ferricyanide decomposition under acidic conditions. Po-tassium ferricyanide ($K_3[Fe(CN)_6]$) was added to concentrated hydrochloric acid and kept at 95 °C. The reaction leads to the formation of a complex mixture containing cyanide coordinated $Fe^{3+/2+}$ compounds (including unreacted potassium ferricyanide) and soluble hexaaquairon (III) ($[Fe(H_2O)_6]^{3+}$) irreversibly obtained in the form of $FeCl_3$ along with hydrogen cyanide (HCN).[52] The conversion of low spin ferricyanide to high spin hexaaquairon (III) is accompanied by an increase in the paramagnetic strength of the reaction mixture. Results presented in Figure 5 depict higher values in the normalized PL contrast two minutes after initiation of the reaction. This result is consummate with an increase in PS concentration equivalent to a conversion efficiency of 60% ferricyanide to $FeCl_3$ (Figure 5A). The following apparent plateau is continued with a slow increase of the contrast possibly due to side reactions with slow kinetics. An apparent slow accumulation of PS is observed in the control solution containing only potassium ferricyanide which undergoes reactions with mirroring kinetics (Figure 5B).

The $MW_{on}$:$MW_{off}$ measurement protocol was also used to map the presence of PS in synthetic liposomes attached to a diamond surface (Figure 6). The $MW_{on}$:$MW_{off}$ protocol was used to produce images of liposomes. Here, 5000 pairs of MW on resonance and MW off images, with 20 ms exposure time, were acquired producing two image stacks. For each stack, single images were generated by summing corresponding pixel values in all images within each stack. The final image was produced by dividing the final $MW_{on}$ and $MW_{off}$ images (Figure 6a and 6c). Differential interference contrast images were also acquired for the same fields of view to demonstrate co-localization of contrast for the $Gd^{3+}$-labelled liposomes with DIC images (Figure 6b) and lack of contrast for the control liposomes (Figure 6d). These findings demonstrate the potential the sensing protocol introduced here has for widefield mapping of endogenous PS produced within the intracellular compartments of live cells.

CONCLUSIONS

A new approach to detection of PS within biological systems is reported. The sensing protocol exploits the spin dependent PL properties of $NV^-$ color defects in diamond and uniquely uses a detection protocol based on successive acquisition of images of PL with and without MWs. Indeed, the experimental system described requires small sample volumes (100 fL), operates at room temperature and requires a relatively straightforward data acquisition and analysis procedure. Varying concentrations of PS in aqueous solutions were detected with a limit of detection for the MRI contrast agent Gadobutrol less than 10 attomol. Exemplar studies of time dependent changes in PS concentration were performed using aliquots of $Gd^{3+}$ salts with increasing concentration and the chemical conversion of low spin cyanide coordinated $Fe^{3+}$ to hexaaqua $Fe^{3+}$ under acidic conditions. Importantly, changes in PS concentration were detected with a time resolution in the range of tens of ms, essential for detecting PS concentration changes associated with short lived bursts of endogenous PS in living systems. Indeed, the endothelium-derived relaxation factor (EDRF), also known as nitric oxide (NO·), is estimated to have a half-life ranging from 0.09 s to more than 2 s depending on the oxygen concentration.[3] Such detection speed is normally very challenging for relaxometry-based protocols which require more than a second per data point.[42,44]

Future work is aimed at elucidation of the mechanism underlying the presented concentration dependent contrast observed in the presence of PS. Previously reported studies of PS using longitudinal spin relaxation of NV color centers[43,4] and recent findings exploiting coupled charge and spin dynamics to detect PS using NV color centers,[53] together with our findings aid in identifying the mechanism driving the phenomena observed here. Indeed, it is known that freely diffusing PS generate magnetic spin noise having a broadband spectral density extending to the GHz range, with frequency components corresponding to the NV Larmor precession. As the concentration of PS increases so too does the NV relaxation rate, which is in part due to the concentration dependent dipole coupling between the PS and NV. This in turn reduces the time dependent probability of finding the NV in the |$m_s$= 0> ground state[42] and hence reduces the polarizability of the NV. It is hypothesized that this would lead to a reduction in both PL intensity and the contrast in PL between MWs on and MWs off contributing to the effects observed in this work. Charge state dynamics in which conversion between the $NV^-$ and neutral NV charge state ($NV^0$) occurs may also factor in the contrast observed. Previous studies report PS affects PL in a concentration dependent manner owing to the interplay between ionization, polarization and recombination of charge states. The role of charge state dynamics in the contrast observed in this study however is not clear particularly as CW optical illumination is used and hence there is no dark time for the suggested recombination mechanism to occur. Whilst recombination may not factor in this work, ionization of the $NV^-$ charge state may be an important factor particularly owing to the CW illumination used. Indeed, ionization of $NV^-$ is known to be a function of optical illumination time[54] and illumination intensity[55].

In spite of this mechanistic uncertainty the protocol presented here extends the current state of the art in NV based sensing and with the use of spin traps to stabilize short lived PS and enable longer acquisition times lower limits of detection are achievable due to improved SNRs. Moreover, this work demonstrates detection of PS in cell models with sub-cellular spatial resolution highly relevant for mapping endogenous PS generation by intracellular compartments. Overall, this work has introduced and demonstrated an important new sensing approach for fast, sensitive imaging of PS in a widefield format, paving the way towards minimally invasive real-time mapping and observation of free radicals in in vitro cellular environments.

# SUPPLEMENTARY INFORMATION

SI Figure 1: ODMR spectra of all solution concentrations studied

SI Figure 2: Graphs of full width at half maximum and resonance locations for varying concentrations of aqueous solutions of PS.

SI Figure 3: ODMR spectra obtained for the excited state transition.

SI Table 1: Data reporting PL contrast as a function of solution pH.

# ACKNOWLEDGEMENTS

The authors wish to acknowledge the European Research Council (ERC) for funding this work through the ERC Consolidator Award, TransPhorm, grant number 23432094. PBW wishes to acknowledge The Royal Society for sup-port through Grant RGS\R1\191154. The authors also wish to acknowledge useful conversations with Dr David Simpson, University of Melbourne Australia.

FIGURES

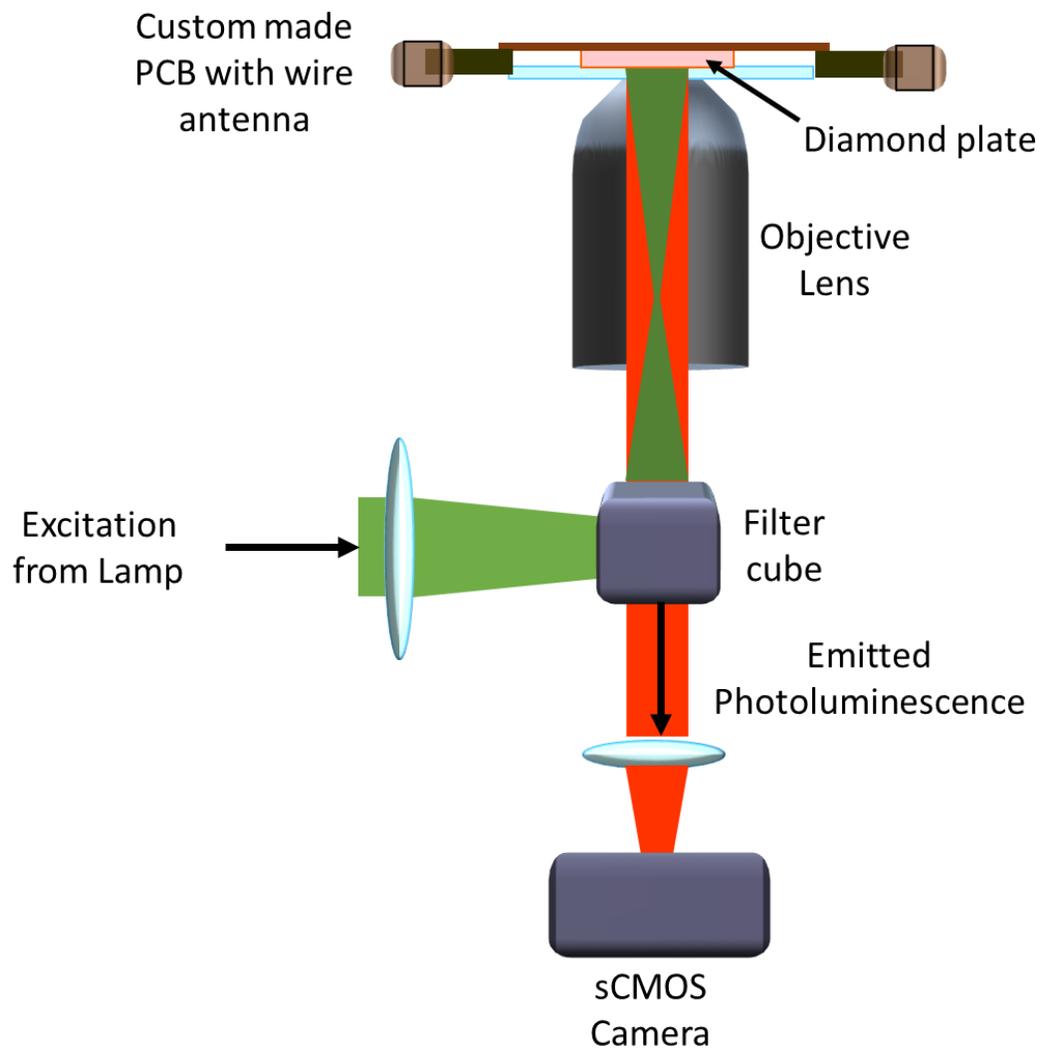

Figure 1: Schematic of experimental set up showing diamond plate on microscope coverslip within custom made PCB with wire antenna and connectors for MW delivery. Optical illumination and detection path also shown.

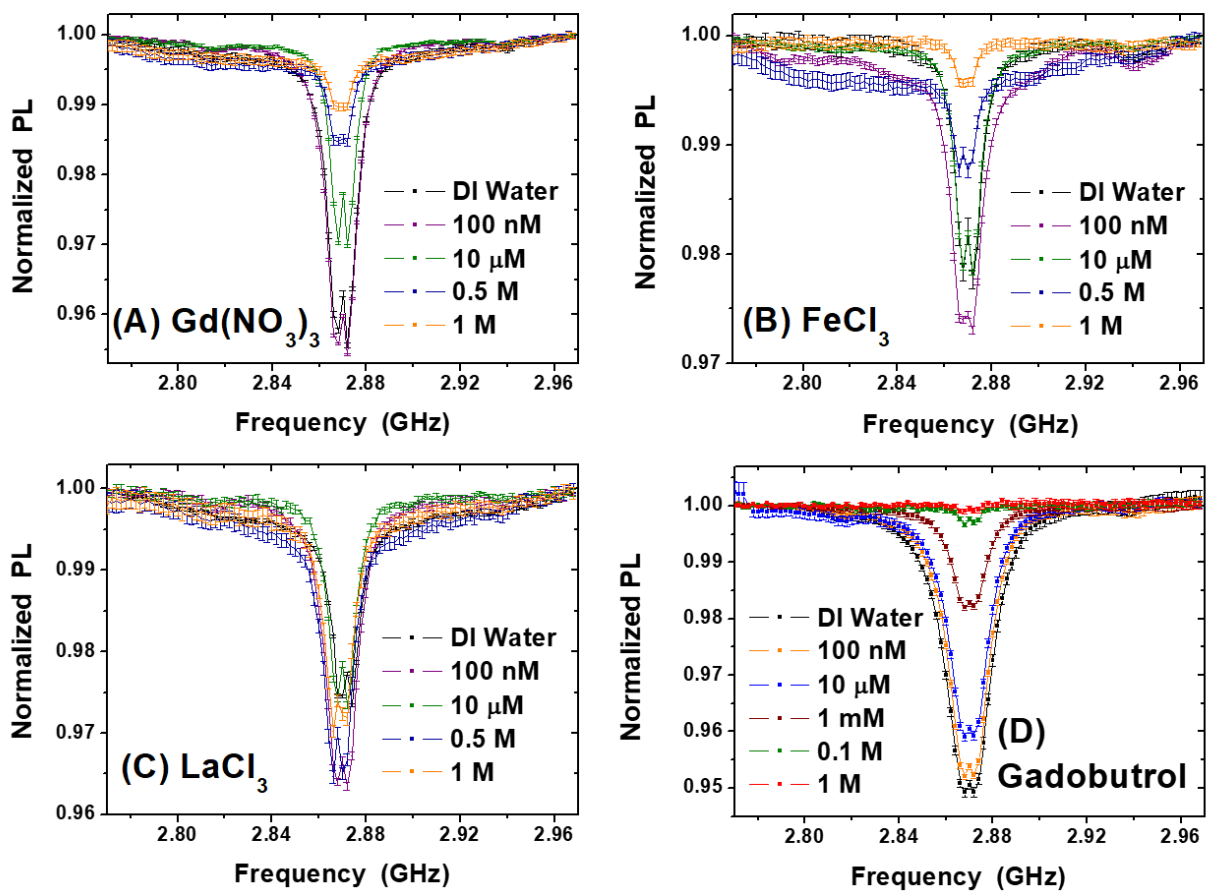

Figure 2: Optically detected magnetic resonance (ODMR) spectra of aqueous solutions of varying concentrations of paramagnetic species: (A) gadolinium nitrate (Gd(NO3)3), (B) ferric chloride (FeCl3), (C) Lanthanum(III) Chloride (LaCl3) and (D) Gadobutrol.

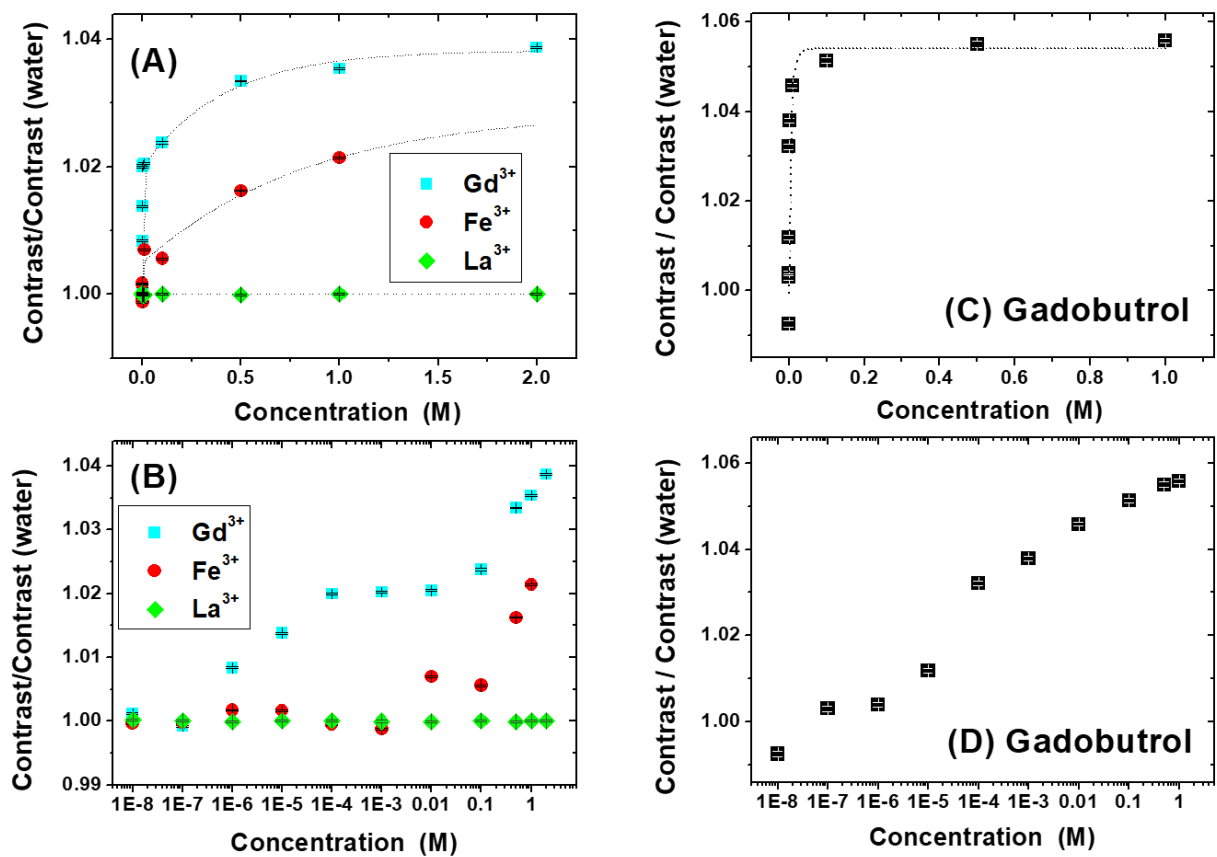

Figure 3: Dependence of the ratio of PL intensity with MW on resonance (2.868 GHz) to PL intensity with MWs off on concentration of aqueous solutions of paramagnetic metals. Data is normalized to results obtained for deionized water. For clarity results are shown on a linear (A), (C) and semilogarithmic scale (B), (D).

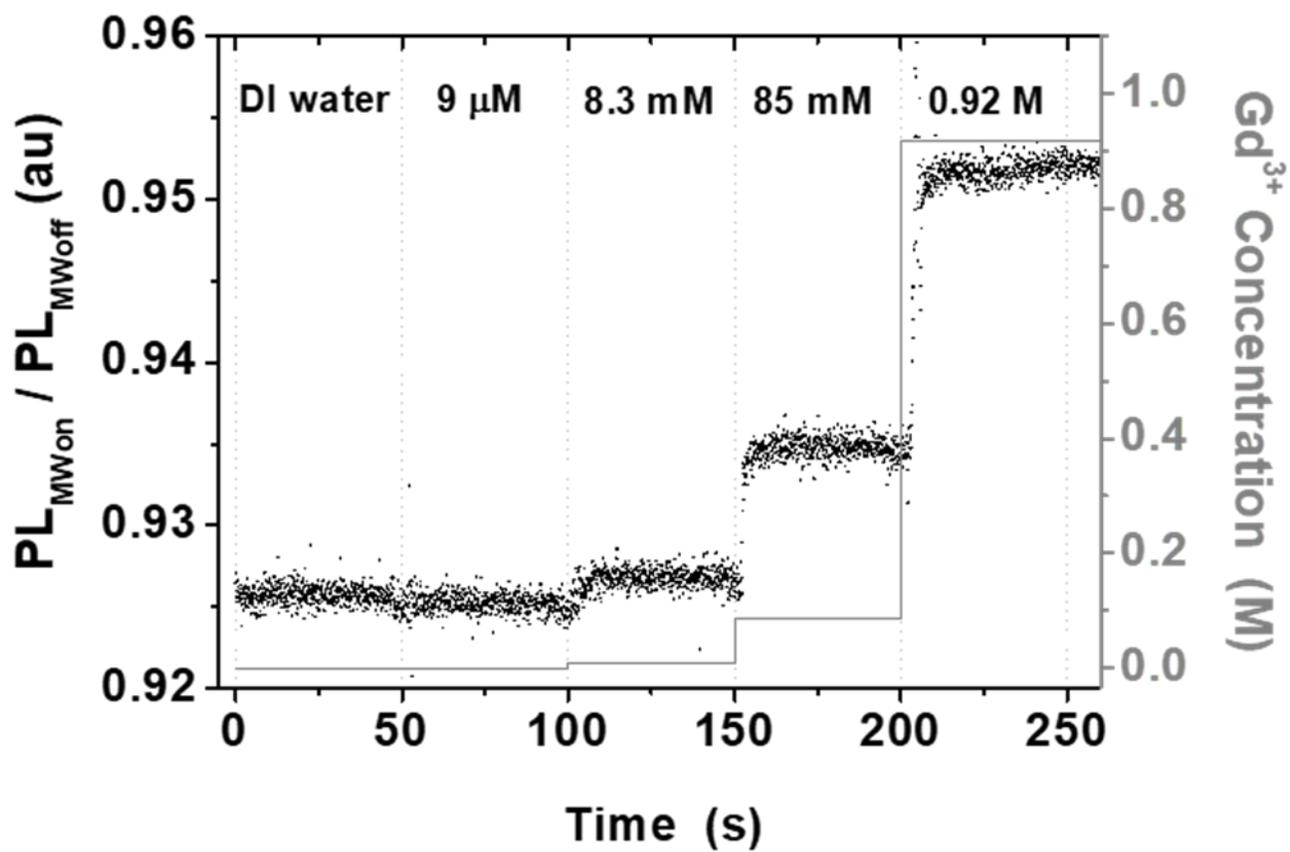

Figure 4: Time course monitoring of contrast obtained from the ratio of PL intensity with MW on (2.868 GHz) to that with MW off (left y-axis) for aqueous solutions with increasing concentration (right y-axis) of Gd(NO3)3.

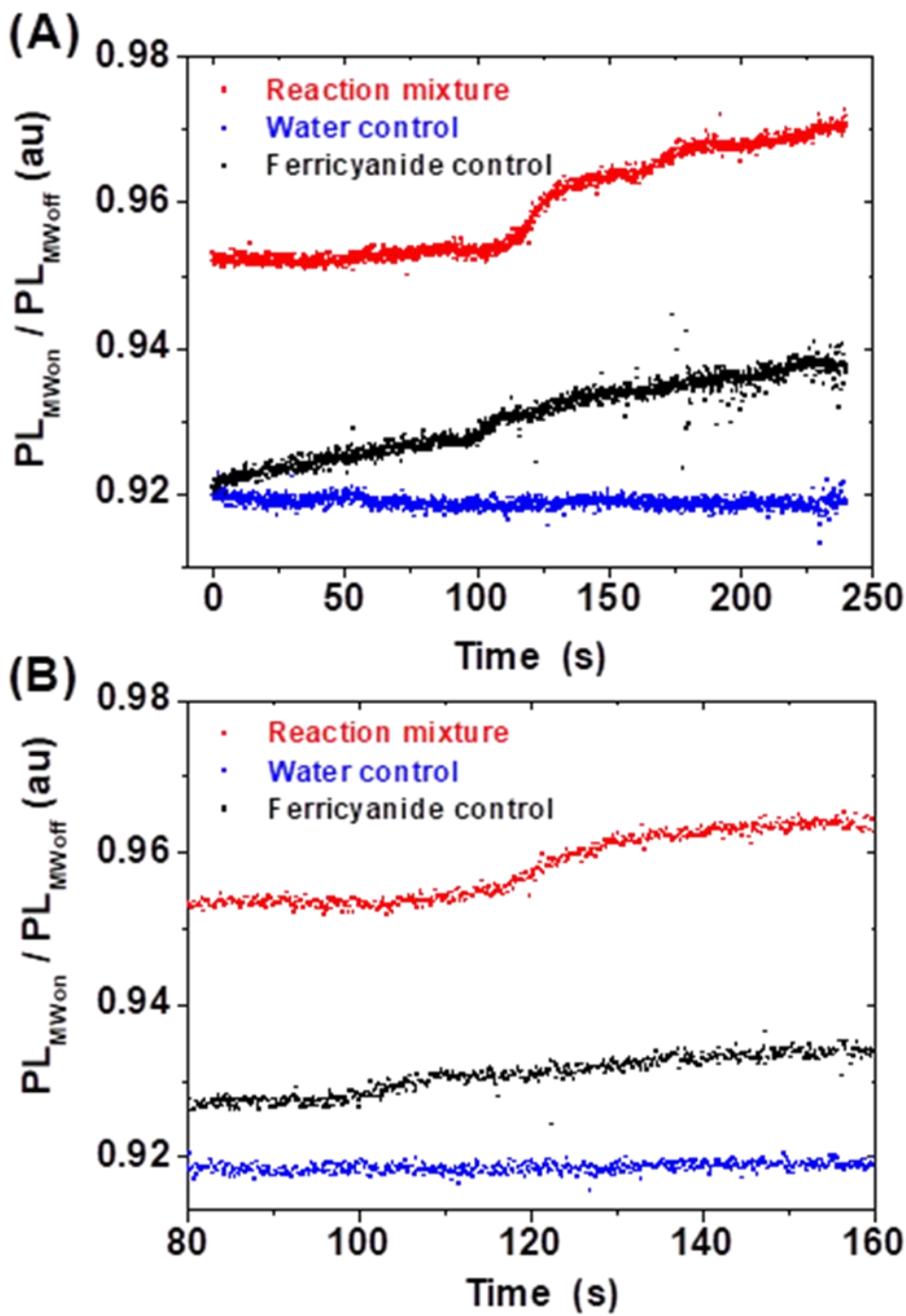

Figure 5: Dynamic study of ferricyanide decomposition under acidic conditions leading to the conversion of low spin ferricyanide to high spin hexaaquairon (III). As the reaction progresses there is an increase in the paramagnetic strength of the reaction mixture. Normalized PL contrast with MWs on resonance and MWs off is shown on the full time scale studied (Figure 5A) and shorter time scale to highlight the early changes in contrast (Figure 5B).

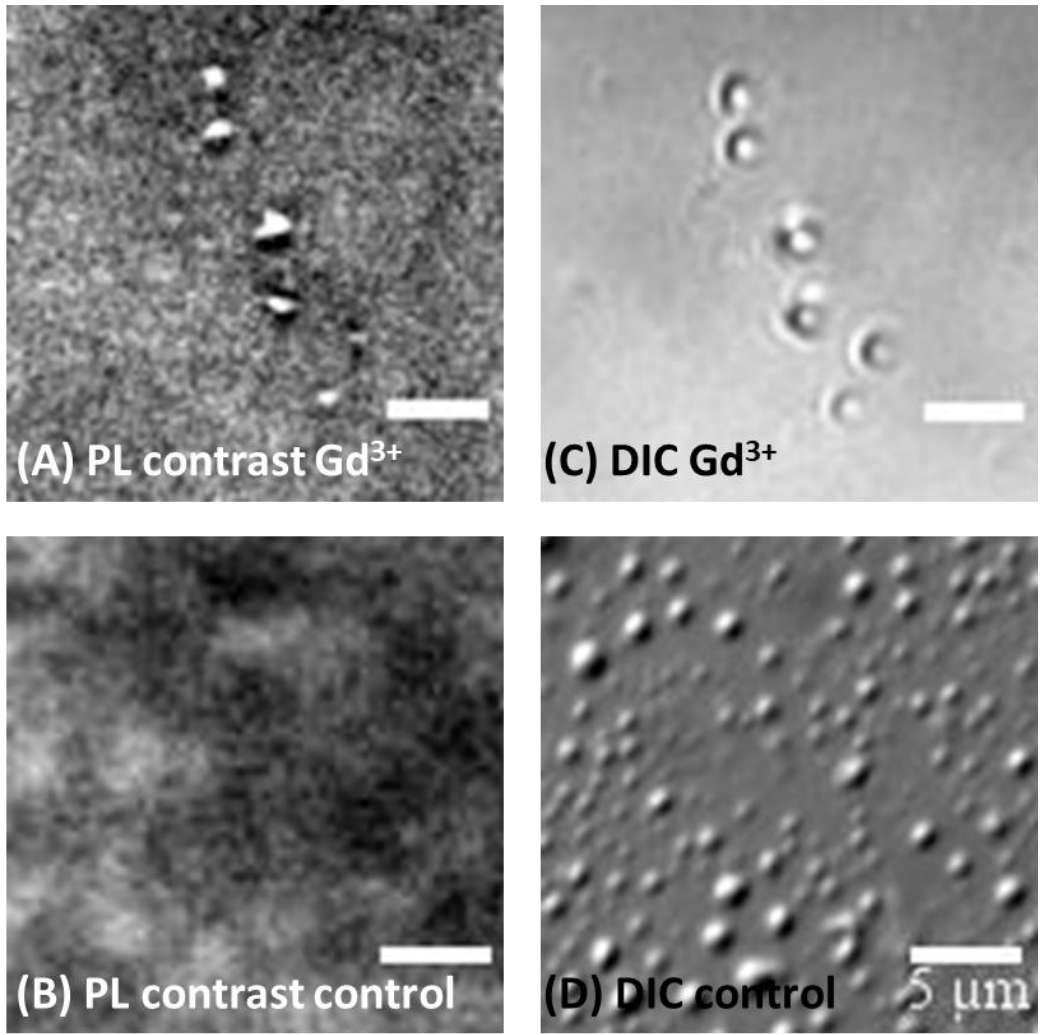

Figure 6: Images showing PL contrast of liposomes containing $Gd^{3+}$ labelled lipids (A) and control unlabeled lipids (B). Corresponding DIC images shown of the same fields of view for the $Gd^{3+}$ labelled (C) and unlabeled liposomes (D).

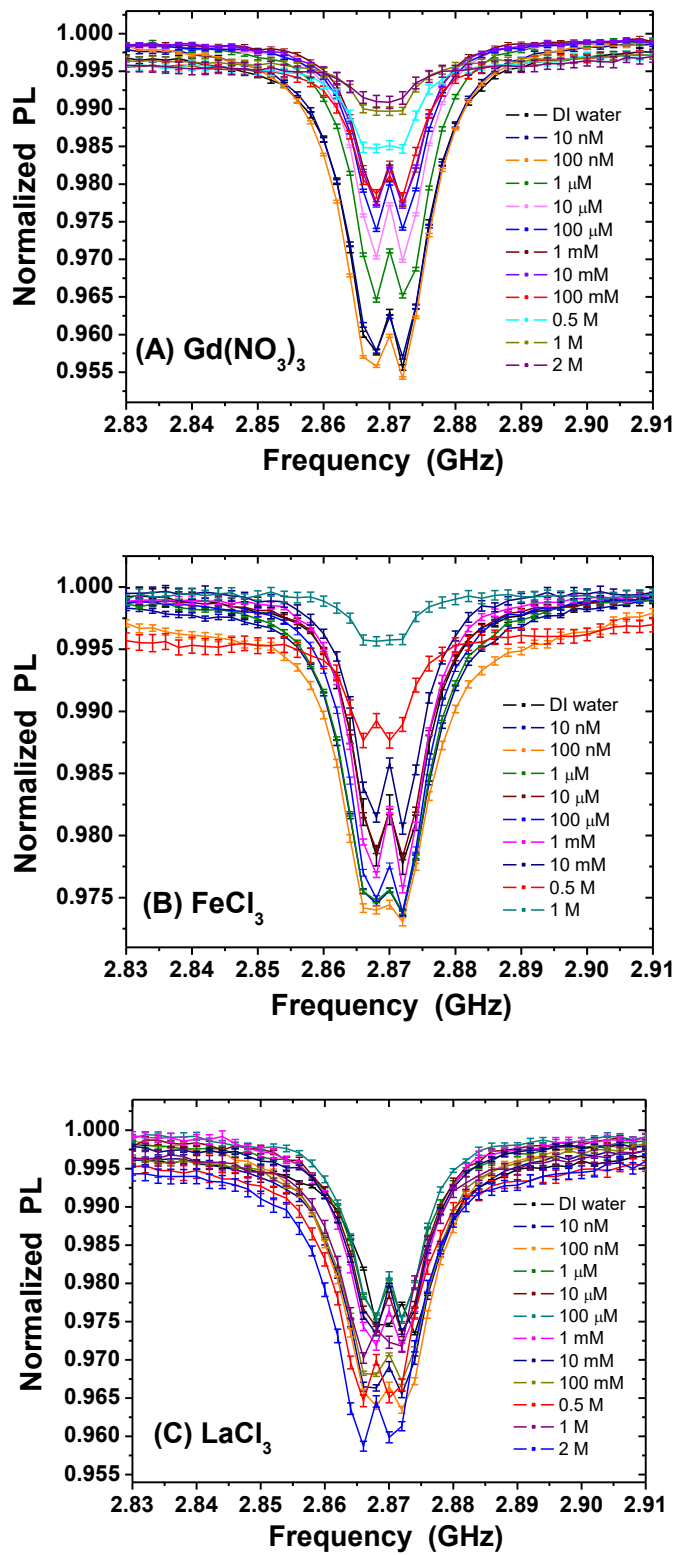

Figure S1: Normalised ODMR spectra of the ground state at varying concentrations of $Gd^{3+}$ (A), $Fe^{3+}$ (B), and $La^{3+}$ (C) in solution. Spectra were recorded from a concentration of 10 nM up to 1 M $Fe^{3+}$ and 2 M $Gd^{3+}/La^{3+}$ (error bars are SEM values).

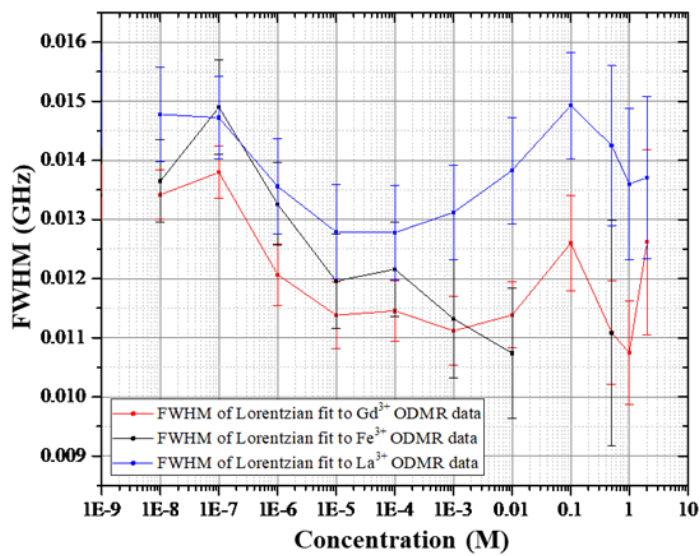

Figure S2: Line width of ODMR spectra as a function of PS concentration for $Gd^{3+}$, $Fe^{3+}$ and $La^{3+}$.

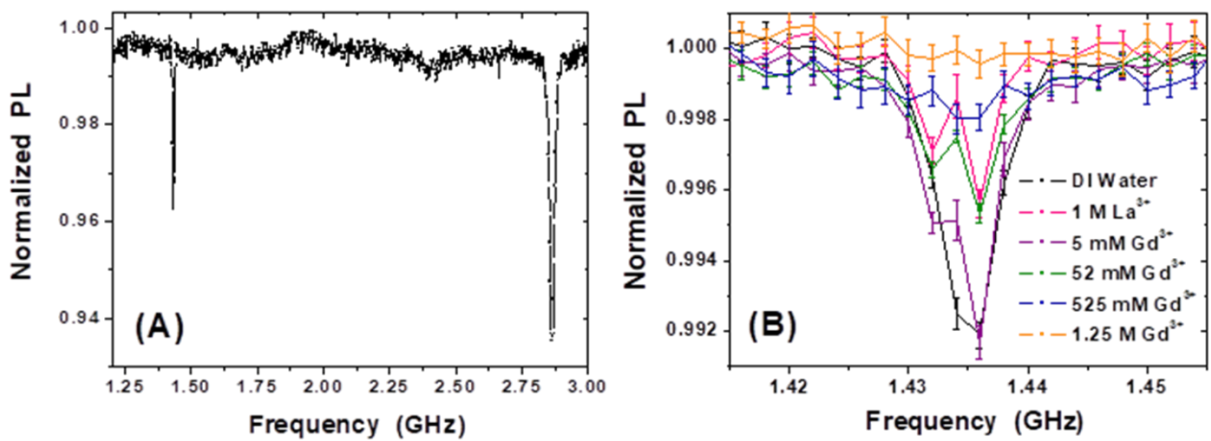

Figure S3: ODMR spectra showing excited state and ground state transitions (a) and excited state transitions for different concentrations of PS (b).

| pH | PL contrast normalized to contrast at pH 7.4 |
|---|---|
| 0.6 | 1.0092 |
| 1 | 1.0080 |
| 3 | 0.9945 |
| 3.6 | 0.9951 |
| 5 | 0.9987 |
| 7.4 | 1.0000 |
| 9 | 0.9971 |
| 11 | 0.9941 |
| 13 | 0.9986 |

Table S1: PL contrast as a function of solution pH normalized to the PL contrast obtained at pH 7.4.